\documentclass[12pt]{article}
\usepackage{bezier}
\usepackage{amssymb}
\usepackage{epsfig}

\title{
Immigration, integration and ghetto formation
}
\author{Hildegard~Meyer-Ortmanns\\
School of Engineering and Science\\
International University Bremen\\
P.O.Box 750561\\
D-28725 Bremen, Germany \\
e-mail: h.ortmanns@iu-bremen.de}

\begin{document}

\maketitle

\begin{abstract}
\setlength{\baselineskip}{1pt}
\noindent
We study ghetto formation in a population with natives and immigrants in the
framework of the two-dimensional Ising-model with Kawasaki-exchange dynamics.
It is the phase structure of the Ising model, the integration speed and the 
immigration rate which determine whether ghetto formation between natives and 
immigrants can be avoided or not. Our simulations are performed in and out-of
equilibrium.
\end{abstract}


\section{Introduction}
Sociophysics as a branch of interdisciplinary research has recently attracted
more attention (for a review see \cite{stauffer1} and for the background
\cite{weidlich1}). Models and tools of statistical physics turned out to be 
successful in reproducing and predicting features of traffic \cite{helbing},
migrations \cite{weidlich2}, and opinion formation in social groups 
\cite{holyst}.

In this paper we study the conditions for ghetto formation \cite{schelling}
in a 
population with natives and immigrants by using Kawasaki-exchange dynamics
\cite{kawa} in
a two-dimensional Ising model. To provide the background from physics we 
summarize some well known features about the phase structure of the Ising model
in section 2. An outline of our measurements that are supposed to imitate
realistic situations is given in section 3. In section 4 we present the
results about ghetto formation. The simulations are parametrized by the 
immigration rate and the integration efforts. Section 5 contains 
the summary and conclusions.


\section{Background from physics}

We study the Ising model in two dimensions with fixed magnetization.  So-called
spins $\sigma$ are associated with the sites of the lattice (here assumed
to be rectangular) taking values of $+1$ or $-1$. These values could stand for
the orientation of the magnetic moment in a ferromagnet, or for the presence 
or absence of an atom in a liquid, or labels for two different metals. In our
application they stand for an immigrant $(+1)$ and a native $(-1)$
in a certain human population. The energy derives from the
interaction between neighboring spins. The energy per pair of nearest neighbors
$x$ and $y$ is given by $-J\cdot \sigma(x)\sigma(y)$, where $J$ is the 
spin-spin interaction. Thus the interaction energy is $-J$ if the spins
are aligned and $+J$ if they are different. When the temperature is high
as compared to $J$, the spins are disordered. However, below a critical 
temperature $T_c$, at zero external field, there is a spontaneous 
magnetization quantified by the non-vanishing order parameter $m=<\sigma>$,
the weighted average over the spin. 
Spins of equal sign like to align and order into a state of broken symmetry. 
For fixed $T<T_c$, at zero
magnetic field $h$, there are coexisting states between
the two different ordered phases with order parameters $\pm m$, respectively.
A plot of $T(\pm m)$ for $T\le T_c$ results in the magnetization curve.
We choose the m-axis as abscissa and the T-axis as ordinate.
Values $(m,T)$ below this curve are forbidden parameters in equilibrium in 
the sense that they lead to unstable or metastable configurations, because
they correspond to values of the magnetization, which would not form 
spontaneously.

If the magnetization is kept fixed, however, as in
the Kawasaki dynamics, $(m,T)$ values below the curve 
lead to phase separation
in the form of infinite-range clusters. In a finite volume, these clusters
have a radius proportional to the linear size of the system. We call these
infinite-range clusters ghettos. Like ghettos these clusters are 
self-supporting in the sense that spins (immigrants) can move over a 
distance of the whole lattice (city, country, their own ``world'') 
without meeting spins of opposite
sign (natives, respectively). Infinite-range clusters should be distinguished
from finite-range correlation volumes with a radius $r \ll L$. Such correlated 
regions correspond to water droplets in the fog with varying 
size, which stay, however, finite, when the system size goes to infinity.
Finite-range clusters will be found as equilibrium configurations for $(m,T)$
values above the magnetization curve.

One way of investigating the phase structure of this model is the 
Monte Carlo method. The ``classical'' way is Glauber dynamics, realized, 
for example, in the Metropolis algorithm. In each step of this algorithm, one
proposes to flip a single spin with an acceptance probability such that
each state occurs with the correct probability. In this approach 
the order parameter 
fluctuates. In the Kawasaki-exchange dynamics, nearest-neighbor spins are 
exchanged under 
heat-bath dynamics, i.e. with probability $1/(1+\exp(\Delta E/k_BT))$, where
$\Delta E$ is the energy change under the spin exchange. This way the
order parameter, here the total magnetization, is conserved during the 
evolution of the system.

In case of the population with natives and immigrants the parameter $k_BT$
plays the role of the so-called social temperature. The social temperature
accounts for the possibility of learning,
assimilation and integration of immigrants
whoever takes the initiative. For simplicity, the population size (total number
of spins) is kept constant during the evolution.
We consider the concentration $c$ of immigrants, $c\in [0,1]$, rather than
the magnetization. Both quantities are related 
via $c=(m+1)/2$, $m\in [-1,+1]$. The magnetization curve will then be
replaced by the coexistence curve. (The more generic name shall indicate
that along and below this curve different phases may coexist.) 

In this dynamics the exchange of spins proceeds only locally in space, while
in reality immigrants will make non-local moves from one place to another. We
do not expect that our qualitative results would change if we skip this
simplification and allow also for exchange with next-to-next nearest 
neighbors, for example, or other local moves. The classical Schelling model 
is roughly a low-temperature simulation of this Kawasaki model but also
allowing for many empty sites \cite{schelling}.

\section {Measurements}
 
We start with a population of $L_1\times L_2$ spins on a square lattice,
initially all chosen to $-1$ (natives). In a first sweep we introduce 
immigrants by flipping randomly chosen $-1$-spins with probability $c$.
The initial population then consists of $(1-c)\cdot L_1\cdot L_2$
natives and $c\cdot L_1\cdot L_2$ immigrants, randomly distributed 
among the natives. The evolution of the system under an exchange of 
natives and immigrants with probability $1/(1+\exp(\Delta E/k_BT))$ is 
then followed over $N$ time steps, i.e. over on average $N$ updates of 
every spin. Observables are measured on the resulting
configuration. In particular we take a snapshot of the configuration as they
are displayed in the various figures below. We then repeat the set of $N$ 
measurements a number of $k$-Monte Carlo step times. Such repetitions serve
to improve the statistics if $N$ was chosen large enough as to give
an equilibrium result. For example, on a lattice of linear size $L$ it takes
about $N\propto L^3$ simulations to produce a ghetto with radius 
$R \sim L$ for a point $(c,T)$ below the coexistence curve. 
If $N\ll L^3$, the k-many
repetitions of the first $N$ sweeps lead to different snapshots, and values 
of the observables differ from the first one by more than statistical
fluctuations. This is just a manifestation of the out-of equilibrium
simulations. 

In realistic immigration and integration procedures the number of 
immigrants grows for a while up to a certain saturation, and integration 
measures are supported with varying intensity. During our runs we therefore
vary the social temperature with speed $\Delta T/\Delta t$ as well as the
concentration of immigrants with a rate $\Delta c/\Delta t$, $\Delta t$
being a time interval in Monte Carlo steps per spin.
Alternatively the final outcome can be parametrized by $\Delta T/\Delta c$,
the slope of the straight line connecting the different measuring points
in the $(c,T)$-diagram. For
each of the selected $(c,T)$-points we perform $k$-times $N$ evolution steps,
where $N$ varies not only with $(c,T)$, but also for fixed $(c,T)$ to study the
in-and off-equilibrium effects.

\section{Results}

We choose a square lattice of size $79\times 100$, where the (unusual) choice
of $79$ is just to facilitate the print of the snapshots of the spin 
configurations (the $\star$ stand for immigrants, the blanks for natives).
We choose periodic boundary conditions. In Fig.1-4 we display the results
of four runs.
\begin{itemize}
\item
Ghetto-dissolution.\\
The first run is at constant immigration concentration of $20\%$ starting 
at a temperature $T/T_c=0.8$ below the concentration curve with $N=10^7$ 
sweeps,
leading to a ghetto of immigrants (Fig.1a). At constant concentration the
integration effort ($T$) is then instantaneously increased to $T/T_c=1.2$
above the $(c,T)$-curve. (Such a quench in temperature may reflect a political
decision to support the early education in the native language and the like.)
The Figs.1b-c (after 100 (1b) and 10000 (1c)
iterations) then show that the ghetto does not dissolve instantaneously. The
time it takes until all remnants of the ghetto have disappeared after
$10^6$ iterations (as shown in Fig.1d) is
comparable to the time as it took to
create it (between $10^6$ and $10^7$ iterations). 
This symmetry is due to the Kawasaki-exchange dynamics and the location in
phase space.
\item
Ghetto formation.\\
Here all $(c,T)$ points are below the concentration curve. For equilibrium
simulations we therefore expect ghetto formation. We start with an initial 
concentration of $15\%$ at $T/T_c=0.8$, simulate little integration efforts 
($\Delta T/\Delta t=0.025/600$), 
although the immigration rate $\Delta c/\Delta t=0.05/600$ is high, 
corresponding to a small slope $\Delta T/\Delta c=0.025/0.05$ 
in the $(c-T)$-diagram. Fig.s 2a and 2b show populations with 
$(c,T/T_c)=(0.15,0.8)$ and (0.25,0.85), respectively, 
after 600 iterations both,
not yet equilibrated, while Fig.s 2c and d show how a ghetto is formed
for $(c,T/T_c)=(0.35,0.9)$ after $N=10^3$ (2c) and $N=10^7$ (2d) iterations.
The equilibrium state (2d) shows a ghetto.
Opposite to the course in 1) it takes time to form a ghetto 
($10^7$ time steps), but it is the unavoidable fate if $\Delta T/\Delta t$
is too small and $\Delta c/\Delta t$ is too large. In other words, politicians
in favor of small integration efforts in spite of a high immigration rate
will be late and encourage ghetto formation.
\item
No ghetto formation at all.\\
These results for $(c,T/T_c)$ points $(0.05,0.8)$ (Fig.3a), $(0.10,0.95)$ 
(Fig.3b), 
$(0.15$ $,1.1)$ (Fig.3c) and $(0.20,1.25)$ (Fig.3d),
all above the $(c-T)$-curve, after $10^7$ iterations certainly in equilibrium, 
shall illustrate that there are situations where ghettos are avoided.
However small the integration efforts are while the immigration slowly 
increases, no ghettos will form. This is what one may have naively expected 
to hold in other regions of the phase diagram as well. It is the well
understood phase structure of the Ising model simulated with Kawasaki 
dynamics that tells us for which $(c,T/T_c)$ values the naive expectation 
fails and for which it holds.
\item
No ghetto formation as an off-equilibrium effect.\\
Here we start with an initial concentration of $15\%$ immigrants at a 
temperature of $T/T_c=0.8$ (Fig.4a) after $N=600$ iterations). 
These are $(c,T/T_c)$-parameters for which the immigrants
would form a ghetto if we wait until the evolution has reached its equilibrium 
state. However, in our simulations we do not wait that long  but
increase $T$ while $c$ increases as well $(0.25,0.85)$ (Fig.4b) and 
$(0.35,0.90)$ (Fig.4c)), both after 600 iterations. Next we limit the
immigration to $35\%$ and further increase the integration measures to 
$T/T_c=1.2$ so that we reach the  
area of phase space above the $(c-T)$-curve where
no ghettos will form. There we
wait $10^7$ iterations 
until the equilibrium situation
has been reached, see the snapshot in Fig.4d. The Fig.'s 4 
demonstrate that no ghettos will 
form whatever our waiting time. The little clusters of immigrants as they
are still seen in Fig.4d correspond 
to ``droplets in the fog'' rather than to an ``ocean
below the atmosphere'' which would be called a ghetto in our context.
In other words: With an appropriate choice of $\Delta T/\Delta t$ and
$\Delta T/\Delta c$ politicians can avoid 
ghetto formation, even if the initial conditions do not look promising
on a long-time scale (equilibrium scale). 
\end{itemize}
\section{Summary and Conclusions}

A general remark is in order about the meaning and value of numbers produced
in our simulations in view of applications to social systems. Certainly none of
the parameter values like the lattice size $L_1\times L_2$, the social 
temperature, the time measured in units of Monte Carlo iterations should be
taken literally. Obviously a realistic typical population size in which
one would study the phenomenon of ghetto formation may exceed our choice of
$79\times 100$ ``spins''. 
Time would be measured in units
of months and years. The social temperature is an effective parameter. It
provides one but certainly not the only possibility to take integration
measures into account. However, what should survive simulations with more
realistic parameters will be the qualitative features summarized in section 4.
An important common feature between social systems and our simulations is
the state of being out-of-equilibrium, at least for intermediate periods. It
was this feature that gave the chance of avoiding ghettos even if the initial
conditions look discouraging under equilibrium conditions.
In summary we have shown that depending on the integration efforts and the 
immigration rate politicians can be both late or in time for avoiding
ghetto formation.

\section{Acknowlegdment}

It is a pleasure to thank Wolfgang Weidlich (Stuttgart) and Dietrich 
Stauffer (K\"oln) for stimulating discussions during the conference on
Sociophysics (June 6-9, 2002) at the Center for Interdisciplinary Research
(ZiF)in Bielefeld.

%
%

\newpage
\begin{figure}[bt]
\unitlength1cm
\begin{minipage}[t]{7.0cm}
\begin{picture}(7.0,9.0)
\put(0,0){\mbox{\psfig{file=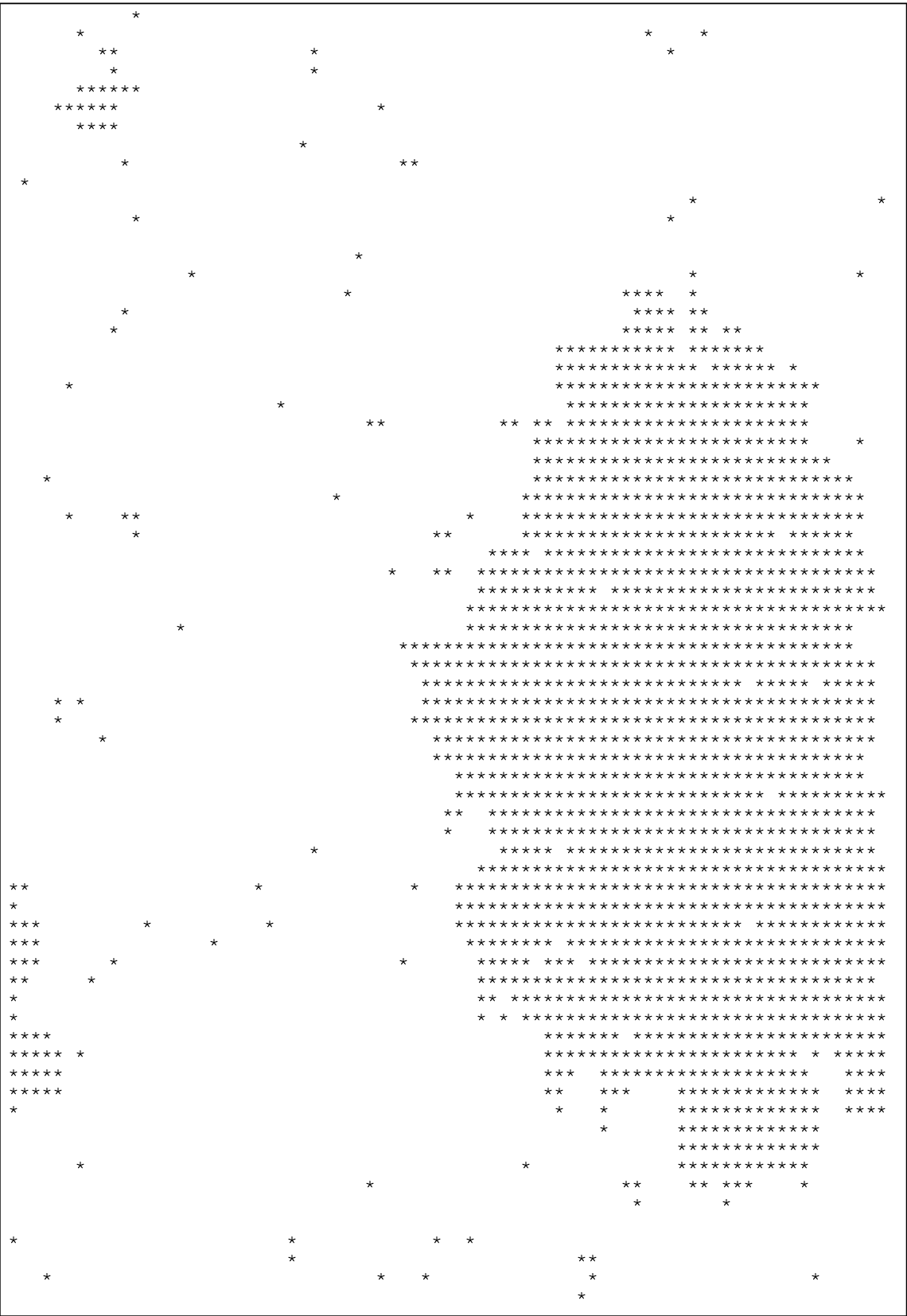,width=7cm,height=9.0cm}}}
\end{picture}
\centerline{ Fig.1a: 
$c=0.2, T/T_c=0.8$,$N=10^7$}
\end{minipage}
\hfill
\begin{minipage}[t]{7.0cm}
\begin{picture}(7.0,9.0)
\put(0,0){\mbox{\psfig{file=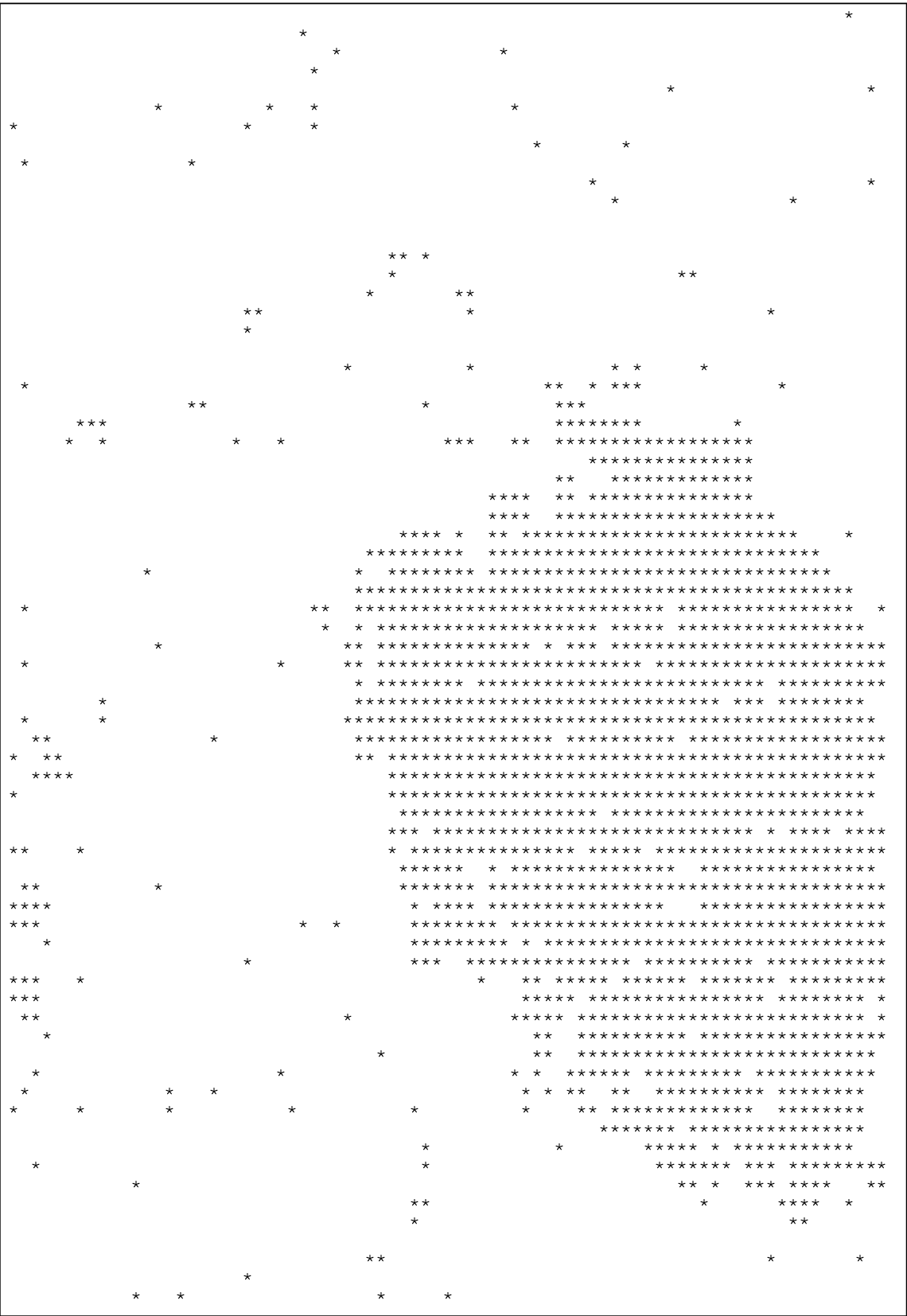,width=7cm,height=9.0cm}}}
\end{picture}
\centerline{Fig.1b: $c=0.2, T/T_c=1.2$,$N=10^2$}
\end{minipage}
\end{figure}
\begin{figure}[bt]
\unitlength1cm
\begin{minipage}[t]{7.0cm}
\begin{picture}(7.0,9.0)
\put(0,0){\mbox{\psfig{file=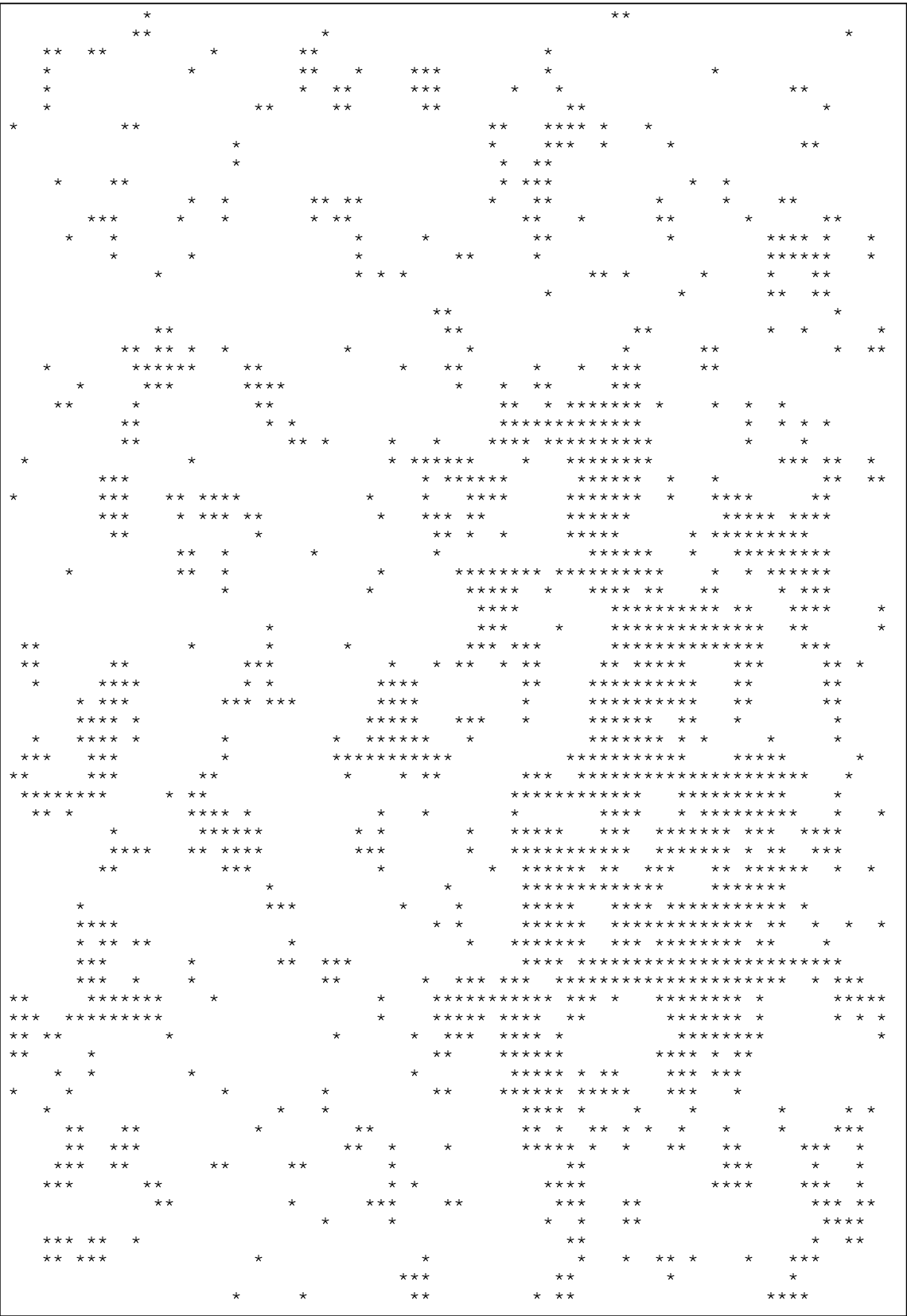,width=7cm,height=9.0cm}}}
\end{picture}
\centerline{Fig.1c: $c=0.2, T/T_c=1.2$,$N=10^4$.}
\end{minipage}
\hfill
\begin{minipage}[t]{7.0cm}
\begin{picture}(7.0,9.0)
\put(0,0){\mbox{\psfig{file=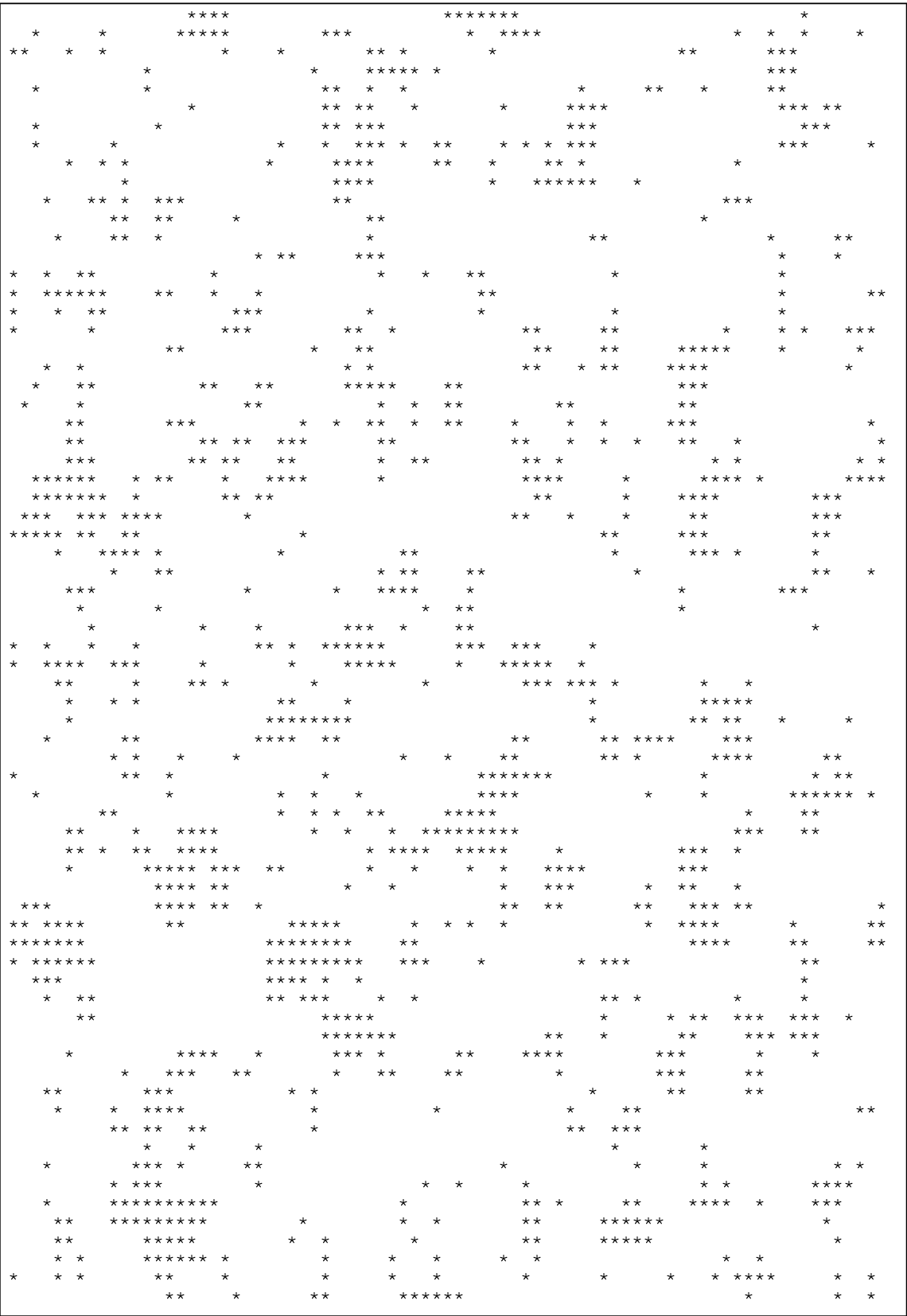,width=7cm,height=9.0cm}}}
\end{picture}
\centerline{Fig.1d: $c=0.2, T/T_c=1.2$, $N=10^6$}
\end{minipage}
\begin{minipage}[t]{15.0cm}
\caption{Dissolution of a ghetto of immigrants $(\star)$ in a population
of natives (blanks).}
\end{minipage}
\end{figure}
\newpage
\begin{figure}[bt]
\unitlength1cm
\begin{minipage}[t]{7.0cm}
\begin{picture}(7.0,9.0)
\put(0,0){\mbox{\psfig{file=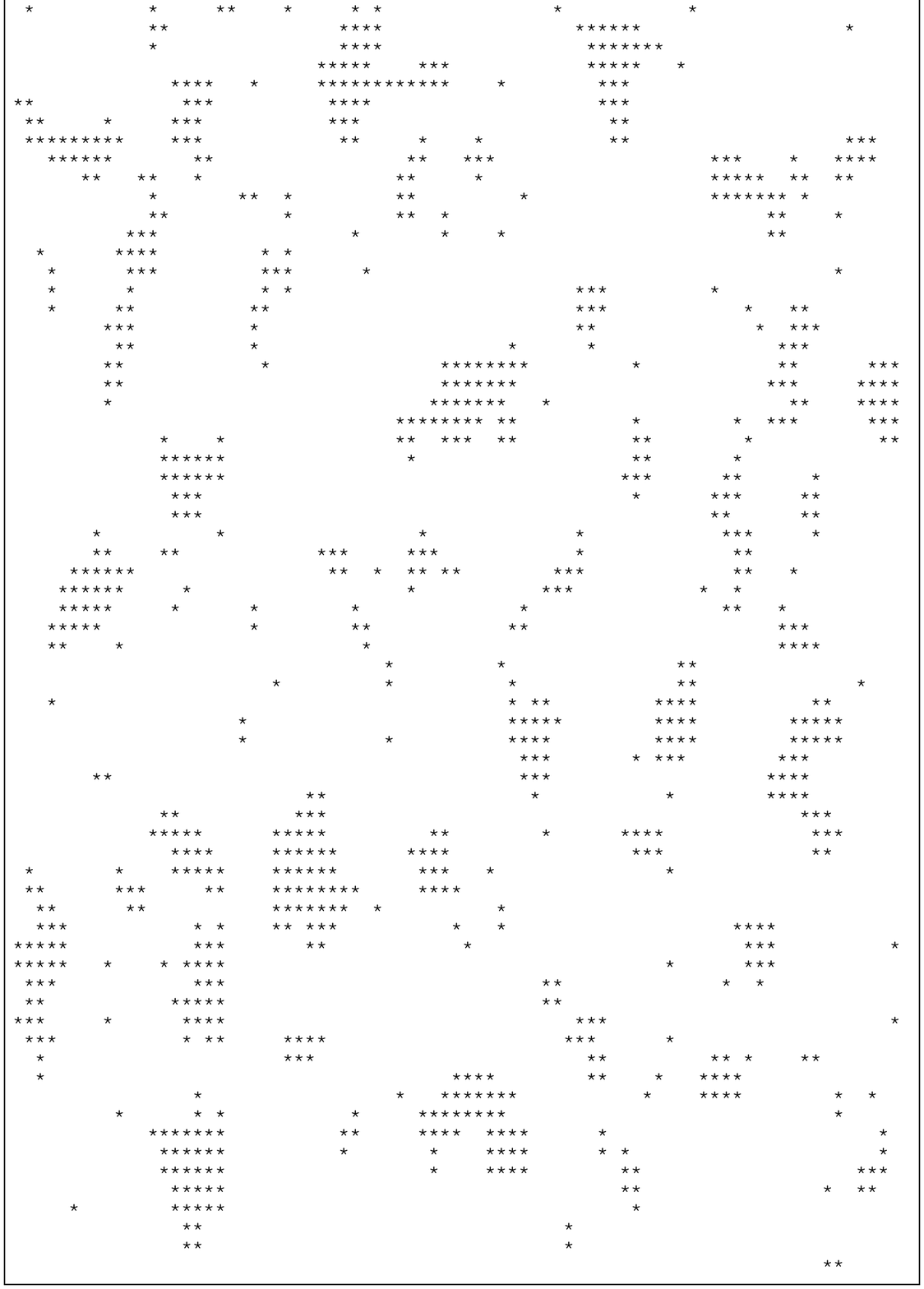,width=7cm,height=9.0cm}}}
\end{picture}
\centerline{Fig.2a: $c=0.15, T/T_c=0.8, N=600$}
\end{minipage}
\hfill
\begin{minipage}[t]{7.0cm}
\begin{picture}(7.0,9.0)
\put(0,0){\mbox{\psfig{file=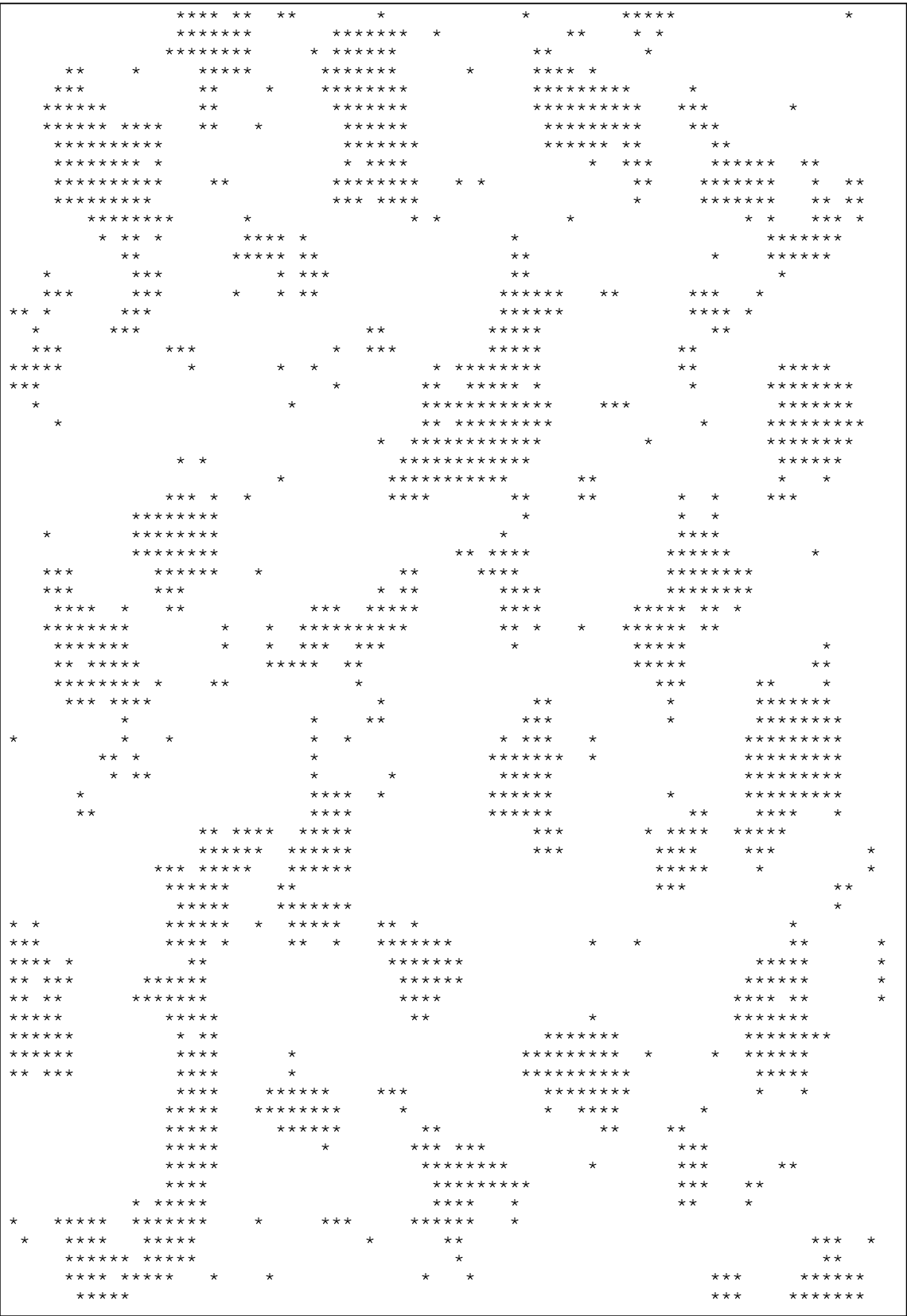,width=7cm,height=9.0cm}}}
\end{picture}
\centerline{Fig.2b: $c=0.25, T/T_c=0.85, N=600$}
\end{minipage}
\end{figure}
\begin{figure}[bt]
\unitlength1cm
\begin{minipage}[t]{7.0cm}
\begin{picture}(7.0,9.0)
\put(0,0){\mbox{\psfig{file=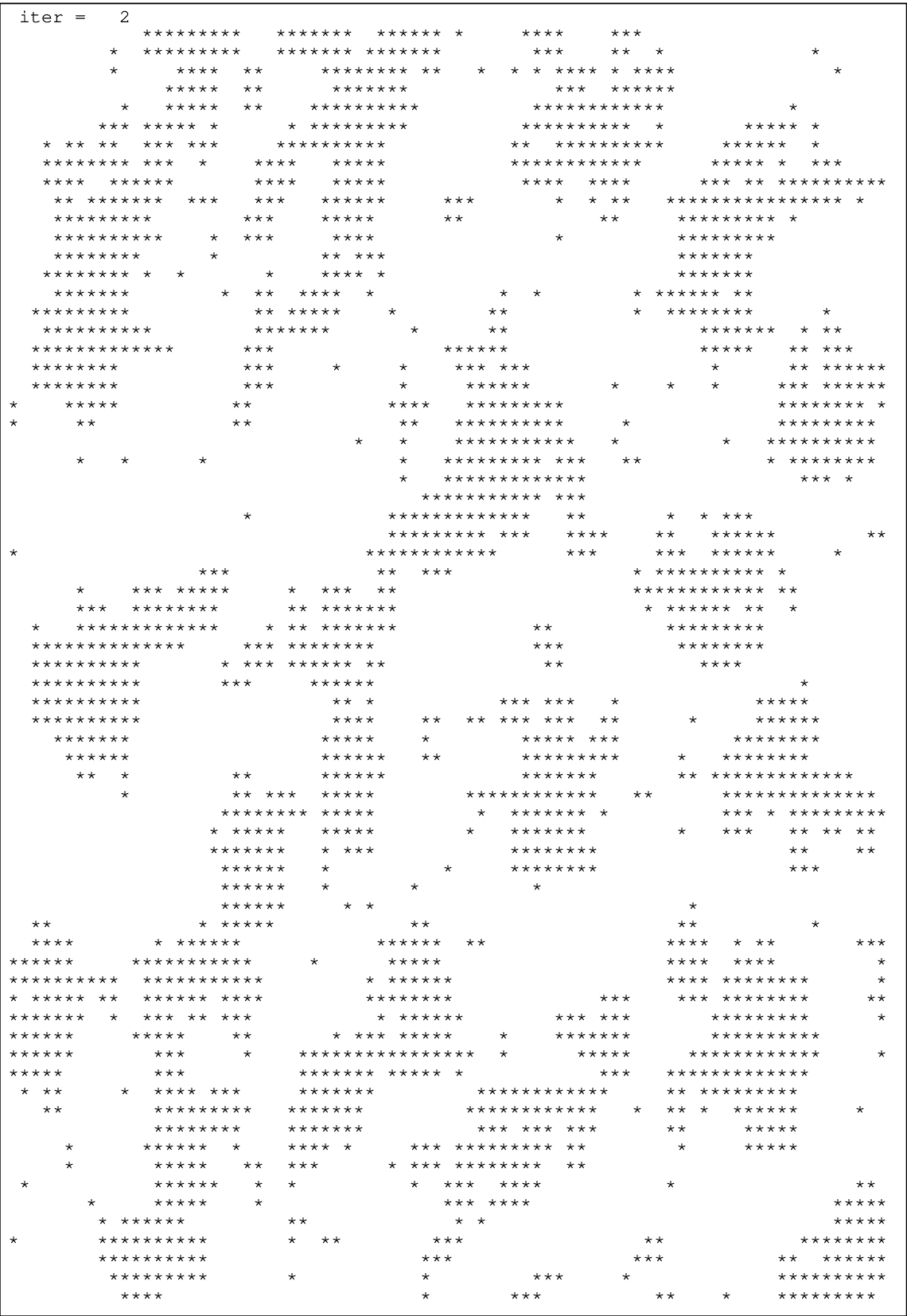,width=7cm,height=9.0cm}}}
\end{picture}
\centerline{Fig.2c: $c=0.35, T/T_c=0.9, N=10^3$}
\end{minipage}
\hfill
\begin{minipage}[t]{7.0cm}
\begin{picture}(7.0,9.0)
\put(0,0){\mbox{\psfig{file=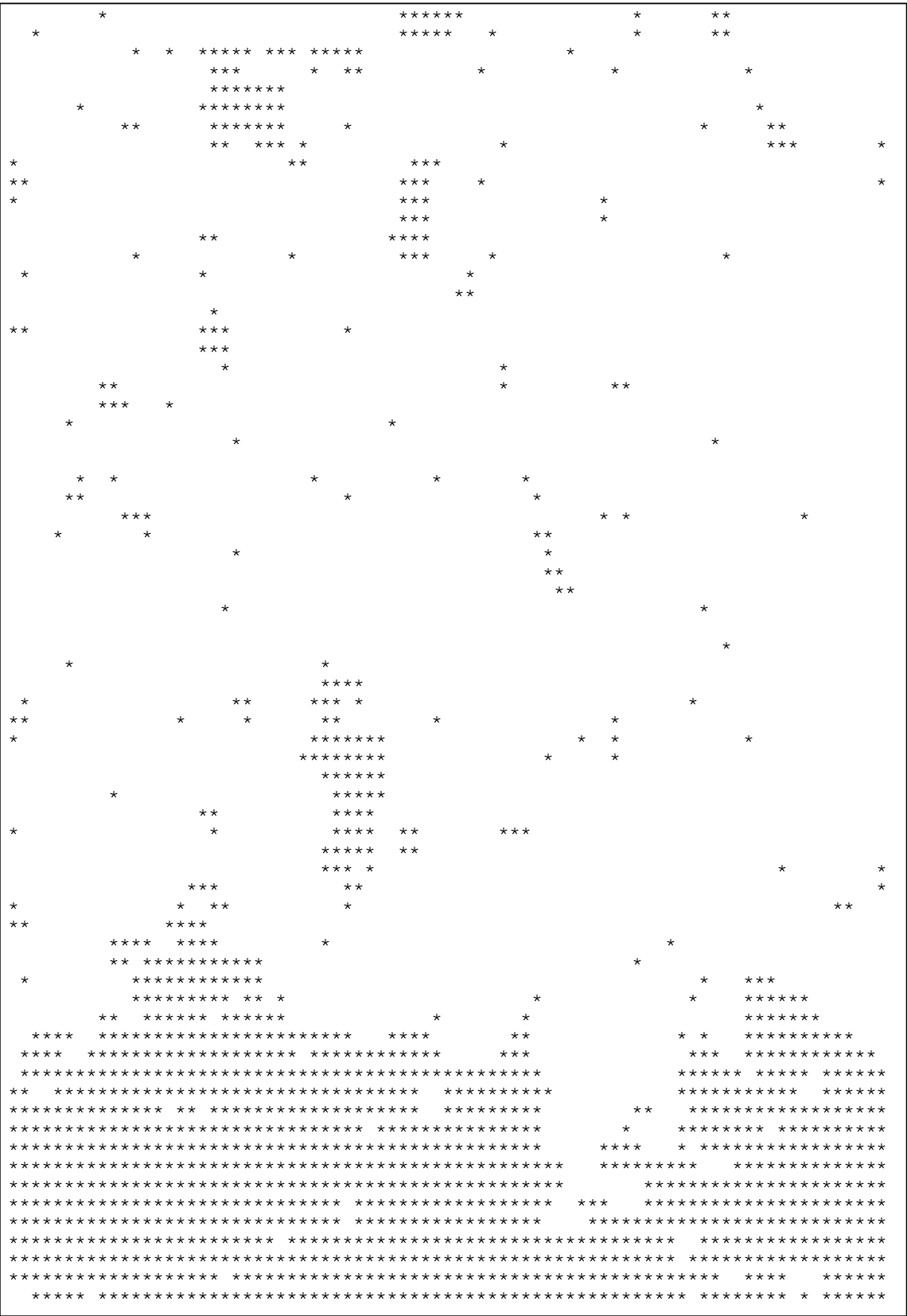,width=7cm,height=9.0cm}}}
\end{picture}
\centerline{Fig.2d: $c=0.35, T/T_c=0.9, N=10^7$}
\end{minipage}
\begin{minipage}[t]{15.0cm}
\caption{Formation of a ghetto.}
\end{minipage}
\end{figure}
\newpage
\begin{figure}[bt]
\unitlength1cm
\begin{minipage}[t]{7.0cm}
\begin{picture}(7.0,9.0)
\put(0,0){\mbox{\psfig{file=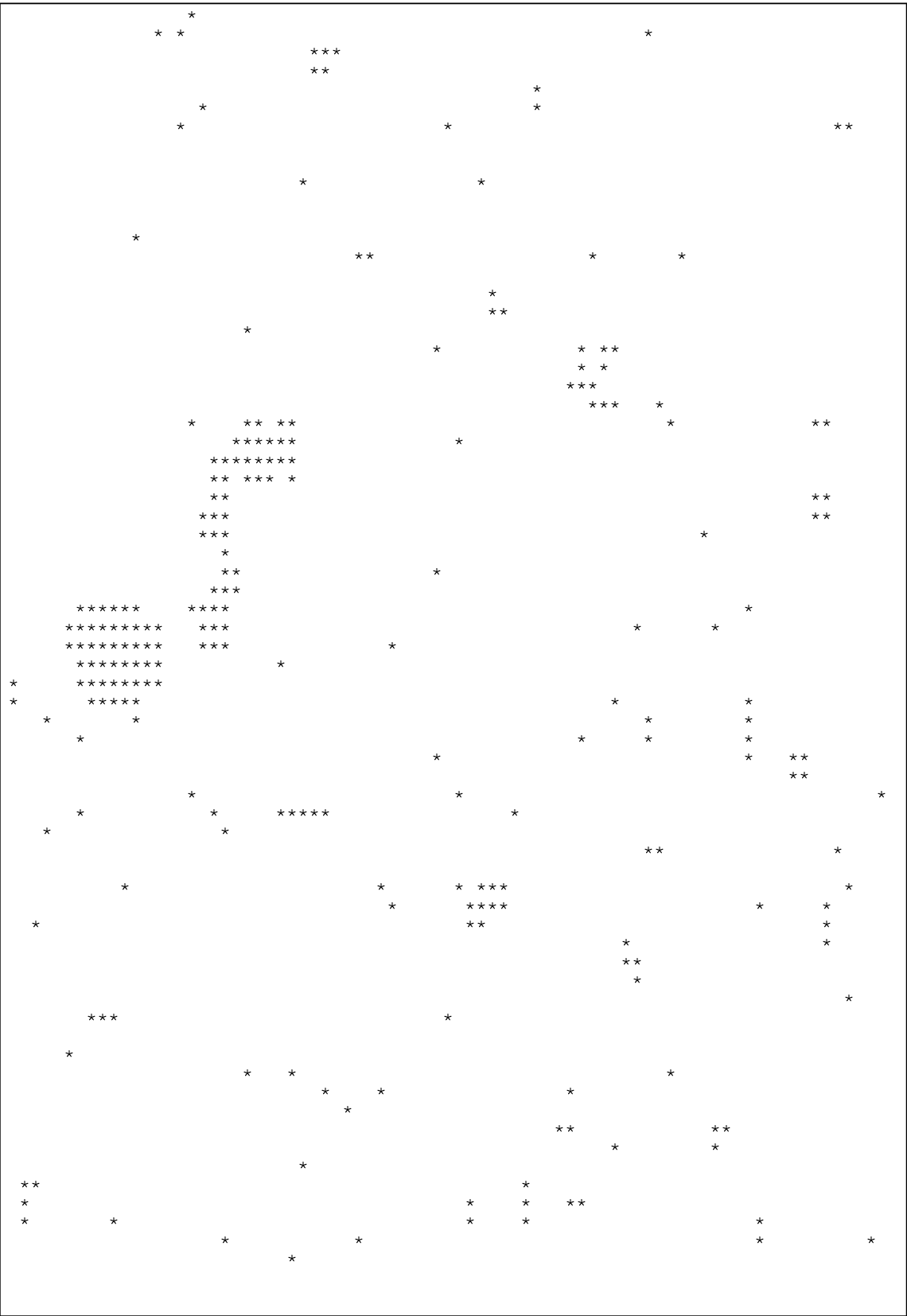,width=7cm,height=9.0cm}}}
\end{picture}
\centerline{Fig.3a: $c=0.05, T/T_c=0.8$}
\end{minipage}
\hfill
\begin{minipage}[t]{7.0cm}
\begin{picture}(7.0,9.0)
\put(0,0){\mbox{\psfig{file=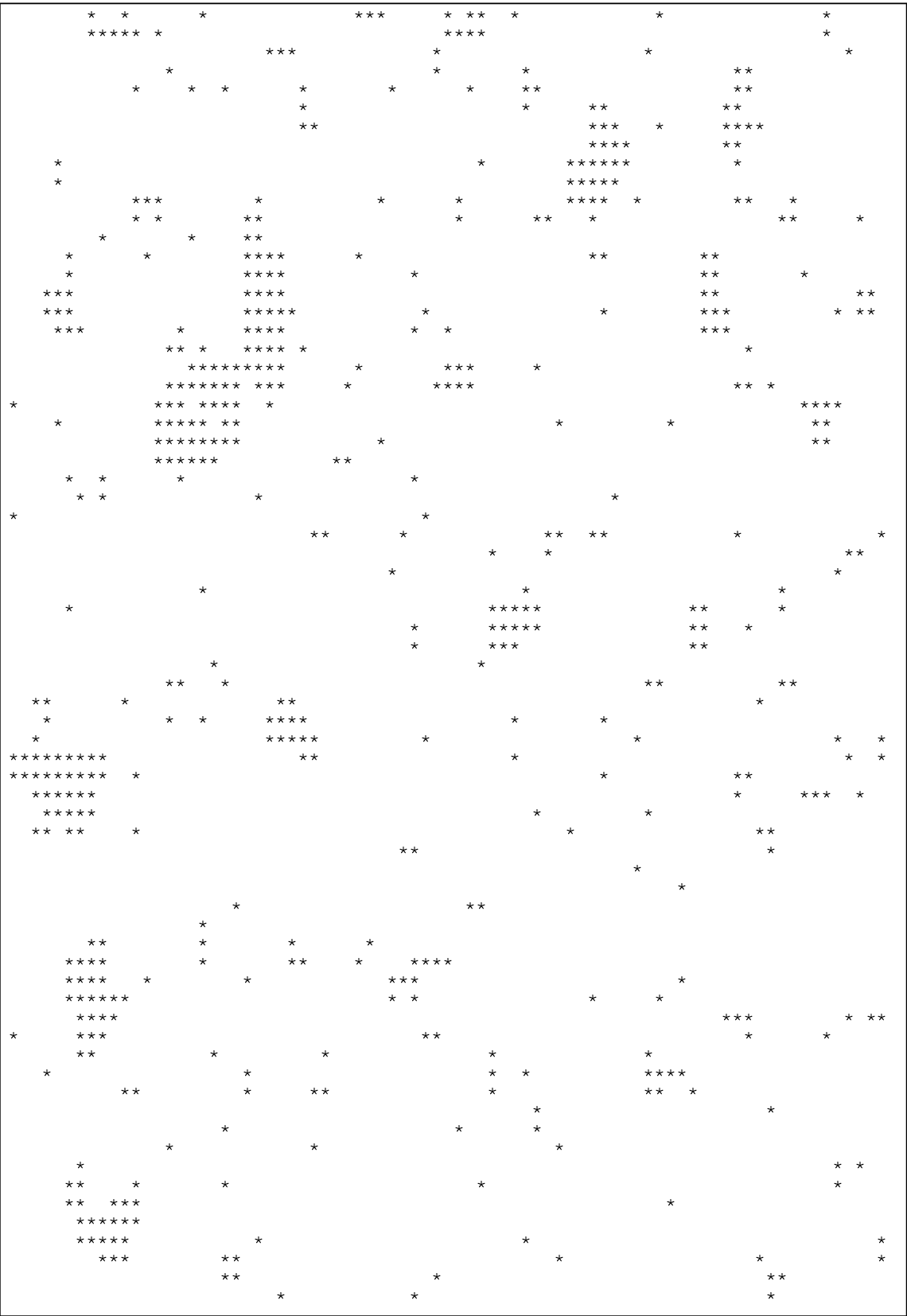,width=7cm,height=9.0cm}}}
\end{picture}
\centerline{Fig.3b: $c=0.10, T/T_c=0.95$}
\end{minipage}
\end{figure}
\begin{figure}[bt]
\unitlength1cm
\begin{minipage}[t]{7.0cm}
\begin{picture}(7.0,9.0)
\put(0,0){\mbox{\psfig{file=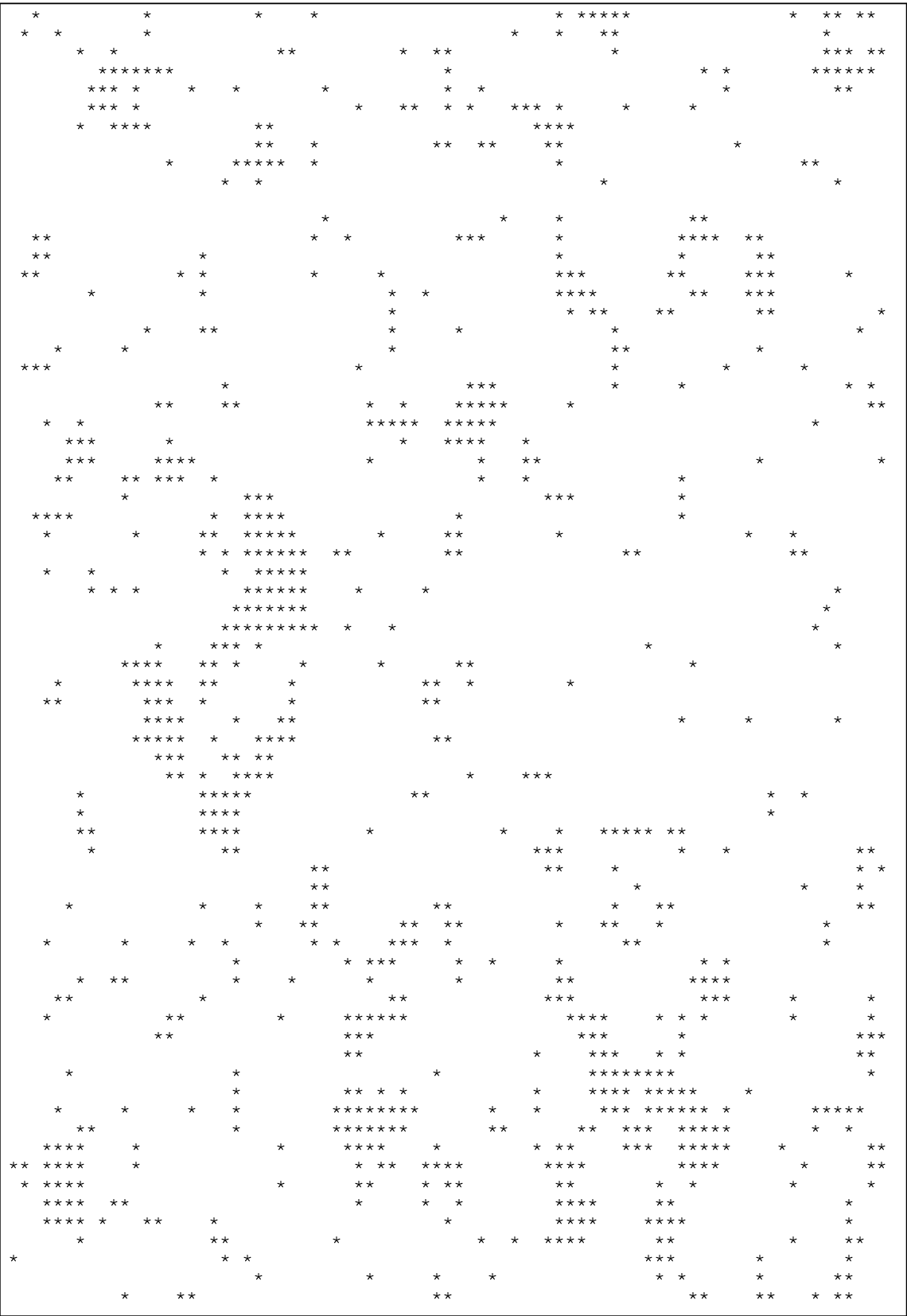,width=7cm,height=9.0cm}}}
\end{picture}
\centerline{Fig.3c: $c=0.15, T/T_c=1.1$}
\end{minipage}
\hfill
\begin{minipage}[t]{7.0cm}
\begin{picture}(7.0,9.0)
\put(0,0){\mbox{\psfig{file=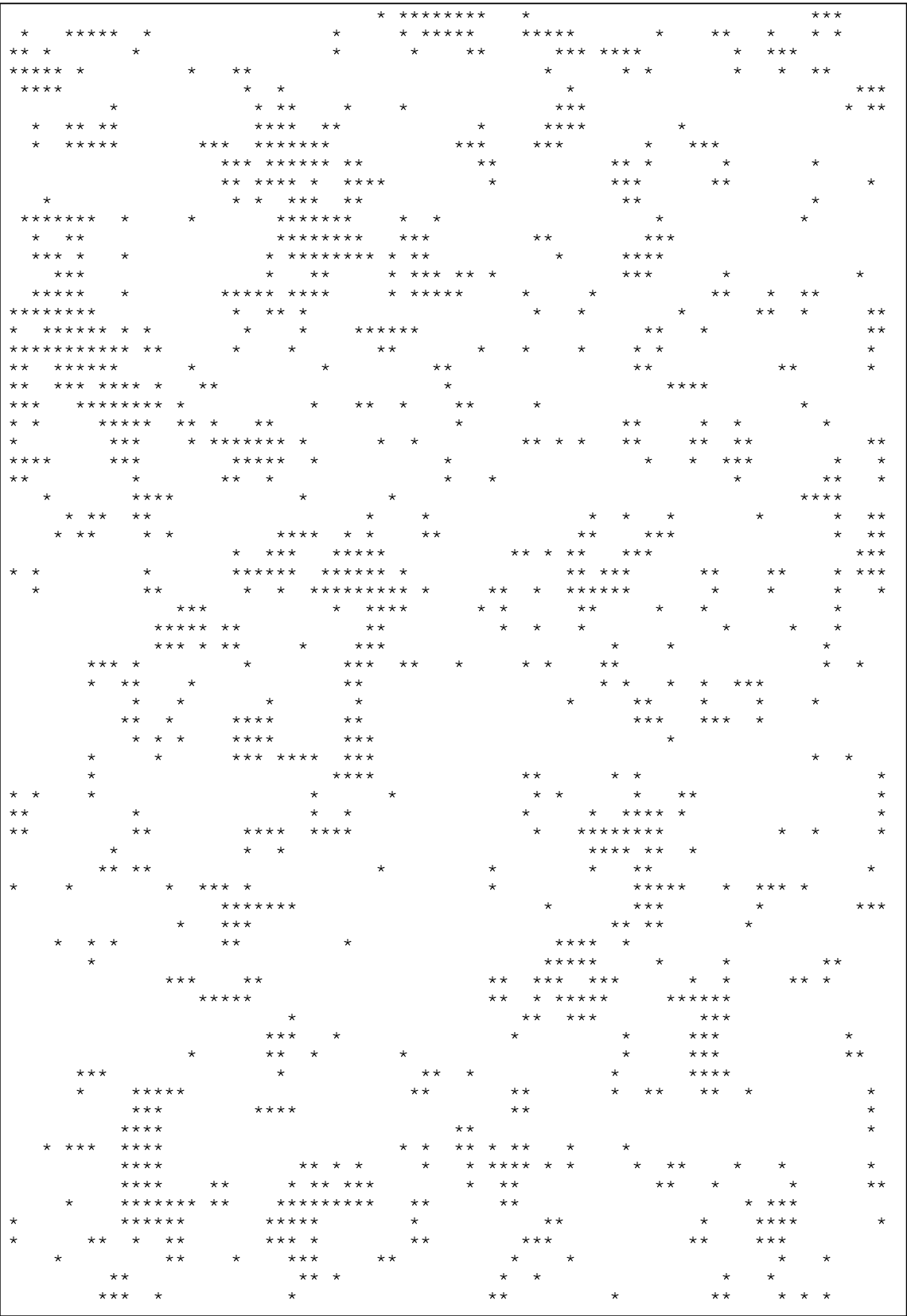,width=7cm,height=9.0cm}}}
\end{picture}
\centerline{Fig.3d: $(c=0.2, T/T_c=1.25)$}
\end{minipage}
\begin{minipage}[t]{15.0cm}
\caption{No ghetto formation in equilibrium. Various concentrations
and temperatures above the coexistence curve after $10^7$ iterations each.}
\end{minipage}
\end{figure}
\newpage
\begin{figure}[bt]
\unitlength1cm
\begin{minipage}[t]{7.0cm}
\begin{picture}(7.0,9.0)
\put(0,0){\mbox{\psfig{file=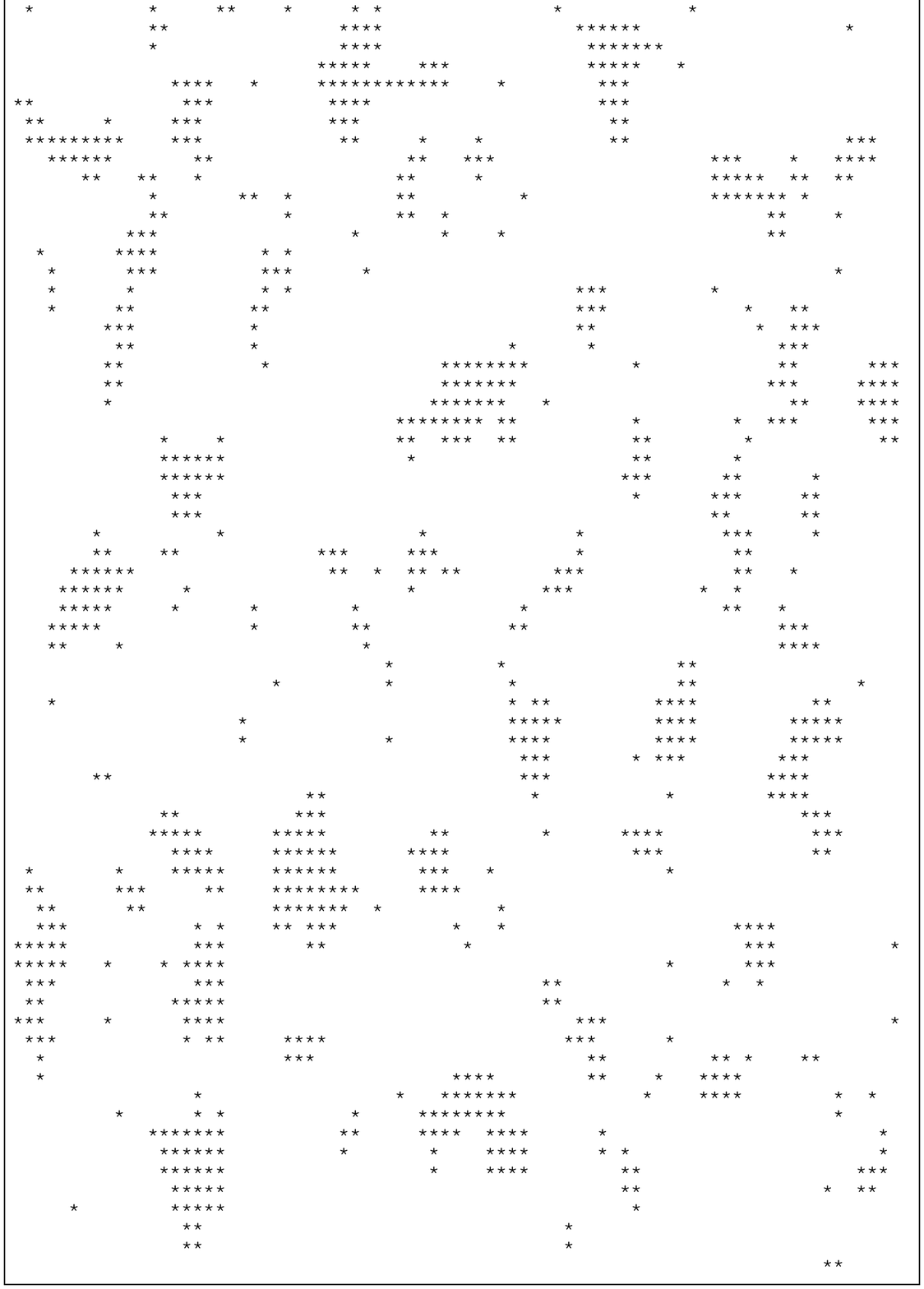,width=7cm,height=9.0cm}}}
\end{picture}
\centerline{Fig.4a: 
$c=0.15,T/T_c=0.8, N=600$}
\end{minipage}
\hfill
\begin{minipage}[t]{7.0cm}
\begin{picture}(7.0,9.0)
\put(0,0){\mbox{\psfig{file=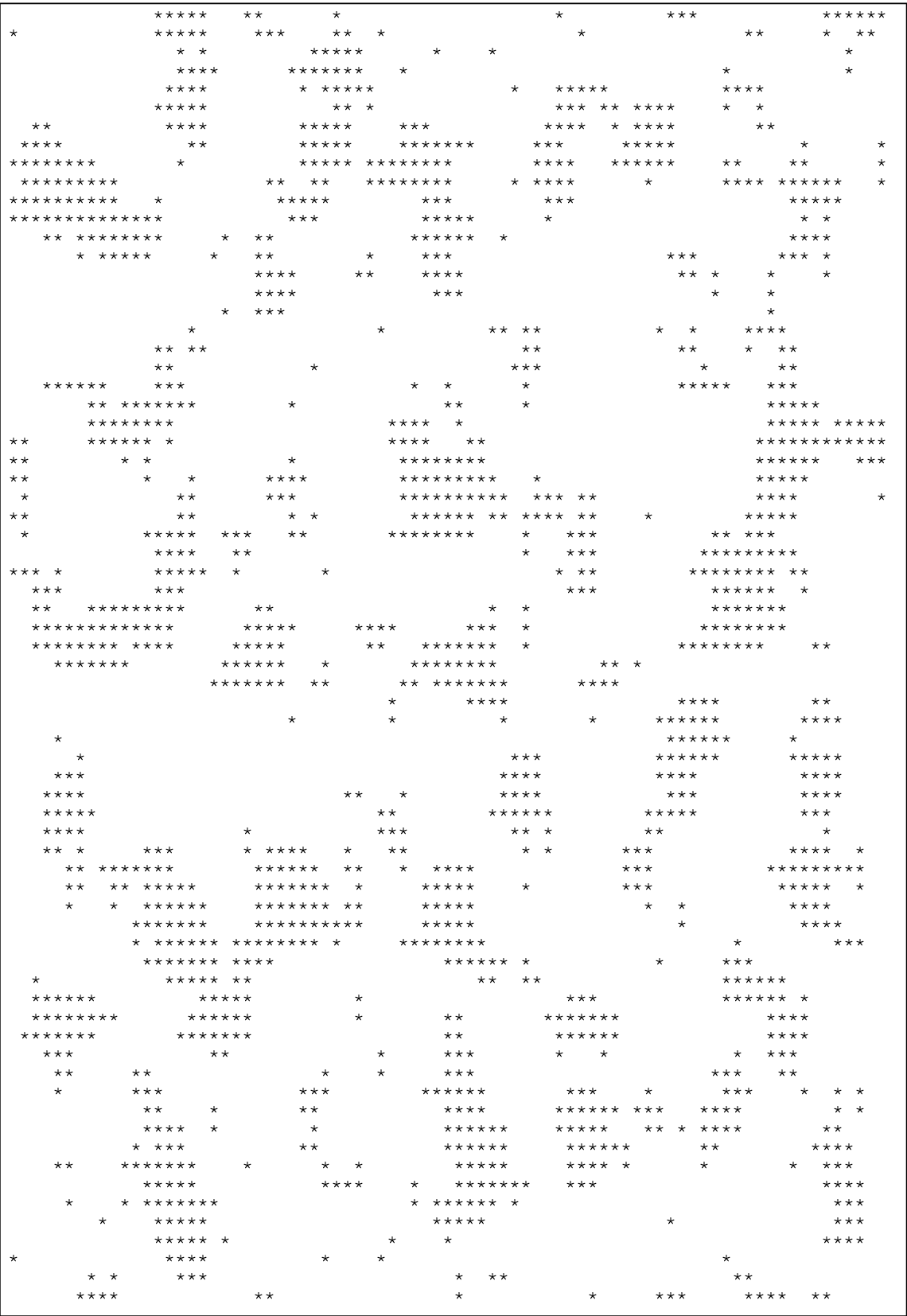,width=7cm,height=9.0cm}}}
\end{picture}
\centerline{Fig.4b: 
$c=0.25,T/T_c=0.85, N=600$}
\end{minipage}
\end{figure}
\begin{figure}[bt]
\unitlength1cm
\begin{minipage}[t]{7.0cm}
\begin{picture}(7.0,9.0)
\put(0,0){\mbox{\psfig{file=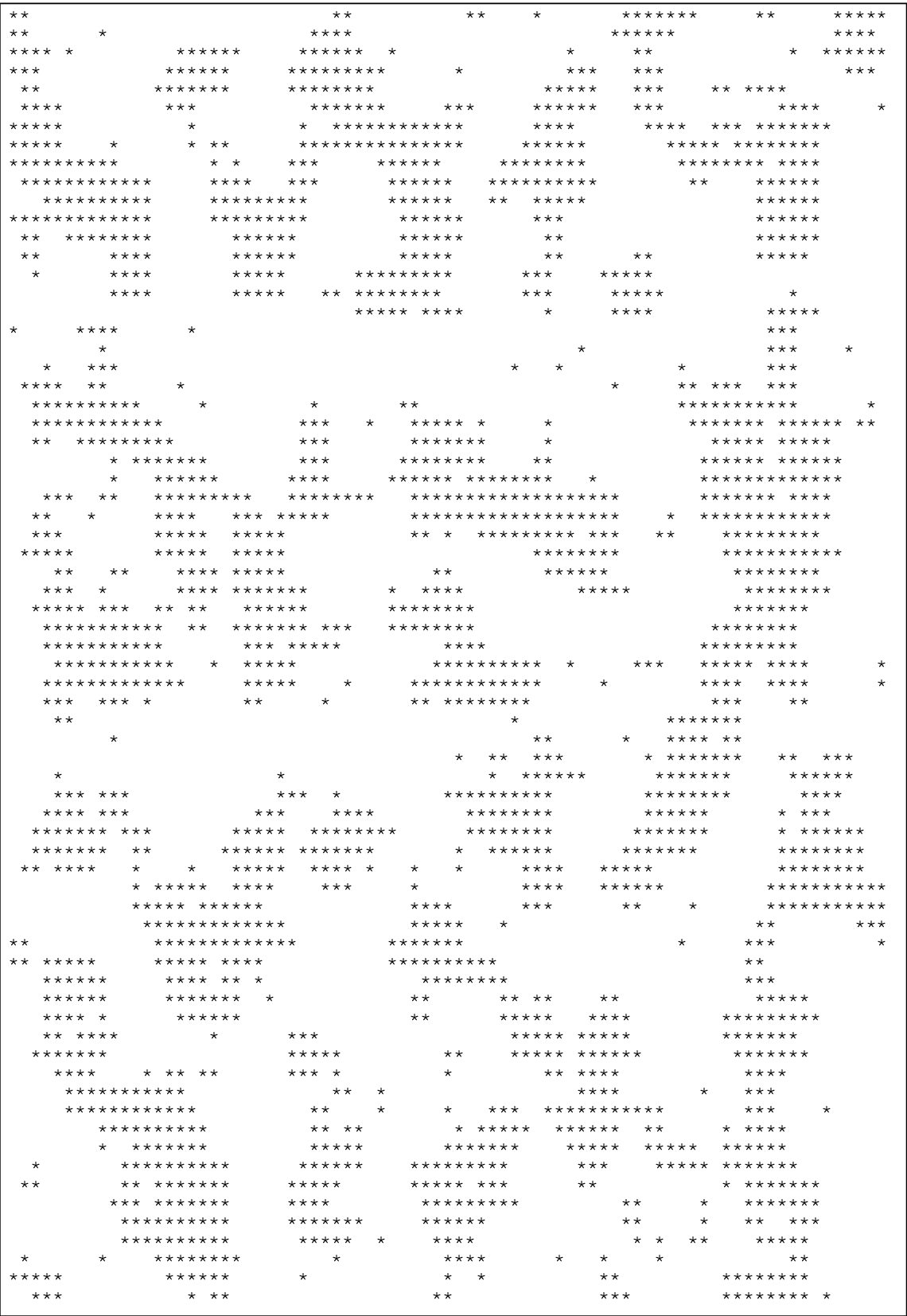,width=7cm,height=9.0cm}}}
\end{picture}
\centerline{Fig.4c: 
$c=0.35,T/T_c=0.9, N=600$}
\end{minipage}
\hfill
\begin{minipage}[t]{7.0cm}
\begin{picture}(7.0,9.0)
\put(0,0){\mbox{\psfig{file=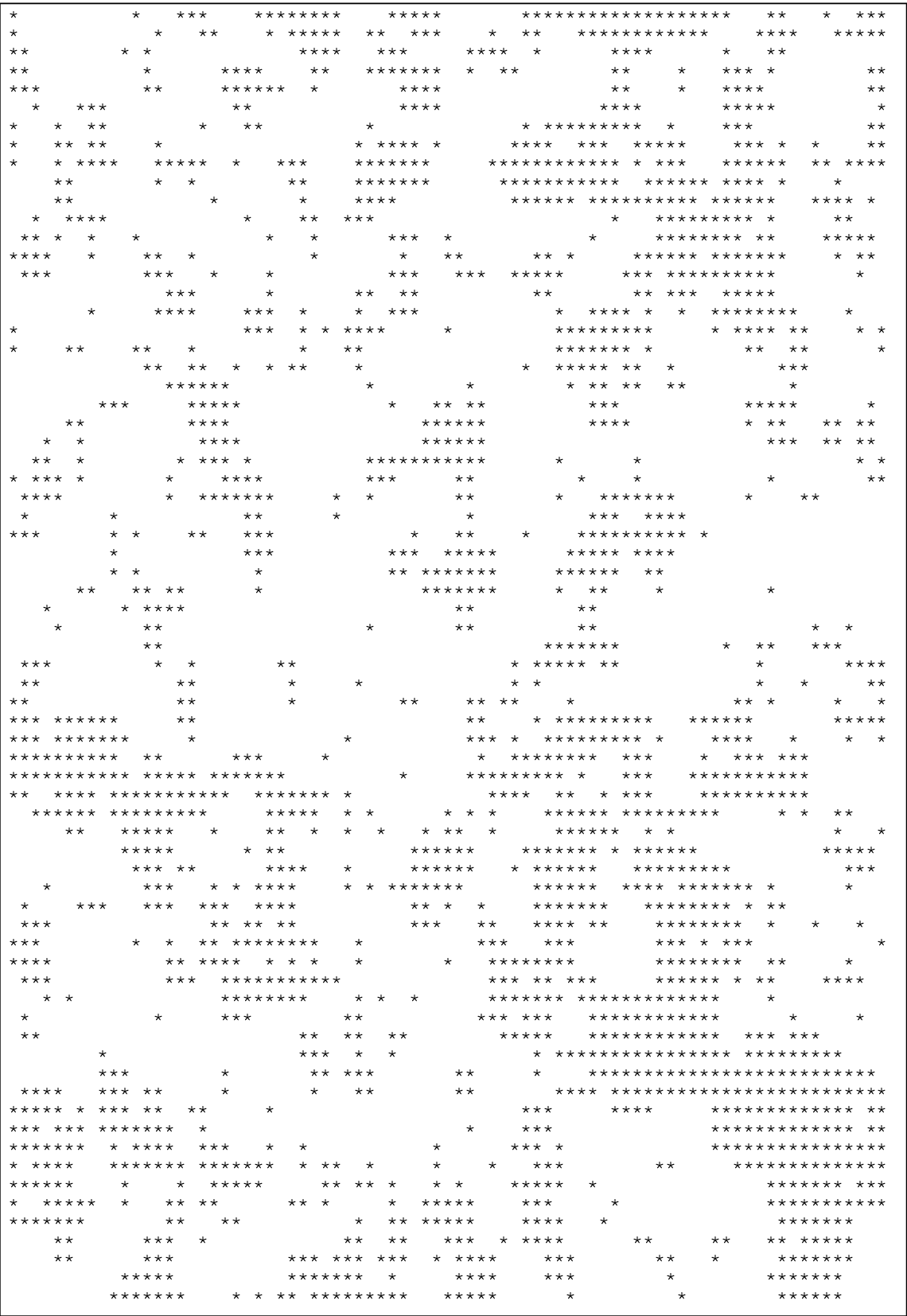,width=7cm,height=9.0cm}}}
\end{picture}
\centerline{Fig.4d: $c=0.35,T/T_c=1.2, N=10^7$}
\end{minipage}
\begin{minipage}[t]{15.0cm}
\caption{No ghetto formation out-of equilibrium below the coexistence
curve (a-c) and none above the coexistence curve (d). }
\end{minipage}
\end{figure}
\end{document}